\documentstyle[12pt,psfig,epsfig,graphicx,psfrag]{article}
\newcommand{\be}{\begin{equation}}
\newcommand{\ee}{\end{equation}}
\newcommand{\ba}{\begin{eqnarray}}
\newcommand{\ea}{\end{eqnarray}}
\def\L5{\tilde{\Lambda}}

\def\N{{\cal N}}
\def\aa{{\cal C}_1}
\def\b{{\cal C}_2}
\def\Mf{M_{(5)}}
\def\MM{M_{(4)}}
\def\a{a}
\begin{document}

\begin{flushright}
hep-th/0311125
\end{flushright}
\vskip 1cm

\begin{center}
{\Large {\bf Solutions of
multigravity theories\\ and discretized brane worlds
}}\\[1cm]
C. Deffayet$^{a,c,}$\footnote{deffayet@iap.fr},
J. Mourad$^{b,c,}$\footnote{mourad@th.u-psud.fr},
\\
$^a${\it GReCO/IAP\footnote{FRE2435 du CNRS.} 98 bis boulevard Arago, 75014 Paris, France.}\\
$^b${\it Laboratoire
de Physique
Th\'eorique\footnote{
UMR8627 du CNRS.
}, B\^at. 210
, Universit\'e
Paris XI, \\ 91405 Orsay Cedex, France.}\\
$^c${\it F\'ed\'eration de recherche APC, Universit\'e  Paris VII,\\
2 place Jussieu - 75251 Paris Cedex 05, France.}
\\
\end{center}
\vskip 0.2cm

\noindent
\begin{center}
{\bf Abstract}
\end{center}
We determine solutions to  5D Einstein gravity
with a discrete fifth dimension.
The properties of the solutions depend on the discretization scheme
we use and some of them have no continuum counterpart.
In particular, we find that the neglect of the lapse field (along the
discretized direction) gives rise to
Randall-Sundrum type metric with a negative tension brane.
However, no brane source is required.
We show that this result is robust under
changes in the  discretization scheme.
The inclusion of the lapse field
gives rise to  solutions whose continuum limit
is gauge fixed by the
discretization scheme.
We find however one particular scheme which leads
to an undetermined lapse reflecting the
reparametrization invariance of the continuum theory.
We also find other solutions, with no continuum counterpart
 with changes in the metric signature or avoidance of singularity.
We show that the models allow a continuous mass spectrum for the
gravitons with an effective 4D interaction at small scales.
We also discuss some cosmological solutions.

\pagebreak

\section{Introduction}

The possibility that space-time has more than four dimensions
has been considered for many years.
The Kaluza-Klein approach,
widely used in superstring constructions and
brane world models, relies on the assumption
that each four dimensional space-time point
is replaced by a compact internal manifold.
Another possibility which
was considered recently is to have an infinite fifth
dimension with a negative cosmological constant, the four
dimensional universe being confined to a 3-brane
\cite{RS} on which gravity is localised.
A generic feature of such models with continuous extra-dimensions is
that the field excitations in the internal manifold are seen by four dimensional observers
as a tower of  infinitely many
massive particles.
This holds in particular for excitations of the metric with non zero
momentum in the internal dimensions, resulting in a tower of massive
gravitons. On the other hand,
theories with a single (or a finite
number of) massive graviton have also been considered.
 In particular it is
well known, that there is only one ghost-free quadratic action for a massive spin 2 field, first introduced
by Pauli and Fierz \cite{Fierz:1939ix}.
Going beyond quadratic order, one is lead to theories with
several coupled metrics \cite{Isham:gm,Boulware:my}.
Such theories, which we will call generically here multigravity,
 have been investigated in the past
in relation to strong interaction \cite{Isham:gm}, and more recently
by various authors \cite{Damour:2002ws,Arkani-Hamed:2002sp} in
particular in relation with
discrete extra-dimensions \cite{Arkani-Hamed:2001ca,Arkani-Hamed:2002sp}.
Discrete extra dimensions have also been considered
in the spirit of non-commutative geometry \cite{Madore:1989ma}.

Theories with massive graviton excitations are
likely to suffer in general from several pathologies which could
prevent them from being very useful to describe the real world. It
includes the celebrated van Dam-Veltman-Zakharov (vDVZ) discontinuity
\cite{vDVZ}\footnote{and the more recently discussed, and not unrelated, possible appearance of
strong coupling
\cite{Arkani-Hamed:2002sp,STRONGDGP}.}, as well as the
propagation of
ghost-like fields when one considers an action which goes beyond
quadratic orders \cite{Boulware:my}. Nonetheless some of these
theories have promising phenomenological properties, in particular in
relation with cosmology \cite{ACCDGP},
 and some of the drawbacks mentioned above seem to be curable
\cite{CUREVDVZ}.
In the perspective of better understanding
theories with multiple
gravitons, it is interesting to contrast theories obtained from a
parent theory which is known to be fully consistent, with the
parent theory itself. This has been done in the past using Kaluza-Klein compactification \cite{Dolan:1983aa}.
Here we follow
a similar path by comparing simple solutions of theories obtained by
discretizing one dimension in five dimensional (5D) general relativity to
solution of the continuum theory.  Namely we  consider the extra dimension to be given by a
one dimensional lattice, to each point of the lattice
is associated a four dimensional space-time
with a metric. We determine the coupling
of the metrics by the requirement that the continuum
limit should be given by five-dimensional gravity.
The discrete action should respect as much as possible
the symmetries of
the 5D action, otherwise new degrees of freedom
appear which are in general ghost-like.
The actions we shall consider break explicitly the
reparametrization invariance along the fifth direction.
Although at the linear level the theory is
free from ghosts, they may appear at the non linear level
\cite{Boulware:my}.

The paper is organized as follows. In Section 2, we  rewrite in a way
convenient for discretization the 5D Einstein-Hilbert action and
 accordingly recall different ways to parametrize 5D AdS space-time as
sliced by 4D Minkowski space-time. We then (Section 3) turn to a first
naive discretization scheme, where the {\it lapse} and {\it shift}
fields (see below)
are omitted. We
show in particular that this allows one to get a brane
world bulk solution
with no source brane included.
However, this solution corresponds to a negative tension brane. We
show next how one can also recover the positive tension configuration,
at the price of introducing a fine-tuned brane source in the
discretized theory. Then we turn to include the lapse in our
discretization scheme, and discuss issues related to reparametrization
invariance along the discretized dimension. In Section 4
we calculate, in the model without lapse field,  the static gravitational potential between two pointlike particles and show that it is 4D at small
length scales and becomes 5D at large distances.
 We then comment on the vDVZ discontinuity in this model.
In section 5 we discuss cosmological solutions of
 the discretized theory.
Our conclusions are collected in Section 6.

\section{Continuum theory: Five dimensional General Relativity}
For the purpose of discretizing a space-like dimension,
parametrized by
coordinate $y$, it is convenient to
 rewrite the 5D Einstein-Hilbert action
\be \label{5DEH}
S_{EH}= \Mf^3 \int d^5 X  \sqrt{-\tilde{g}}  (\tilde{R}-2\L5),
\ee
using a 4+1 splitting of space time.
Above, and in the following, we use  expressions
with a {\it tilde} for
quantities of the 5D continuum theory, like the 5D metric
$\tilde{g}_{AB}$, and upper case Latin letters from the beginning of
the alphabet, $A,B,C...$ to denote 5D indices, $\tilde{\Lambda}$ is the 5D cosmological constant and $\Mf$ is the 5D reduced Planck mass.
After an integration by parts,
(\ref{5DEH}) is rephrased into
\be  \label{ADM} \Mf^3
\int d^4x dy \sqrt{-g} {\cal N} \left\{R -2 \tilde{\Lambda} + K_{\mu \nu} K_{\alpha \beta}\left( g^{\mu \nu} g^{\alpha \beta} - g^{\mu \alpha} g^{\nu \beta}\right) \right\},
\ee
where $K_{\mu \nu}$ is the extrinsic curvature of
surfaces ${\cal H}_y$ located at constant $y$, and we have introduced in a standard
way\footnote{note however that  the surface located at constant $y$
  are timelike, unlike
  in  the usual ADM splitting.} the {\it lapse}
${\cal N}$, the
{\it shift} $N_\mu$, and induced metric $g_{\mu \nu}$ on ${\cal H}_y$
whose Ricci scalar is denoted by $R$.
Note that here and in the following, lower case Greek letters, $\mu,
\nu, \alpha, ...$ will be denoting 4D indices, i.e. indices parallel
to ${\cal H}_y$.  The extrinsic curvature is defined by
\be \label{DEFK}
K_{\mu \nu} = \frac{1}{2 {\cal N}}\left(g'_{\mu \nu}
 - D_\mu N_\nu - D_\nu N_\mu \right),
\ee
where $D_\mu$ is the covariant derivative associated with the induced
metric $g_{\mu \nu}$ and a prime denotes an ordinary derivative with
respect to $y$. The fields ${\cal N}$, ${ N}_\mu$  and $g_{\mu \nu}$
are simply related to the
components of the 5D metric $\tilde{g}_{AB}$
by
\ba
\tilde{g}_{\mu \nu} &=& g_{\mu \nu}, \\
\tilde{g}_{\mu y} &=& N_\mu \equiv g_{\mu \alpha} N^\alpha,\\
\tilde{g}_{yy} &=& {\cal N}^2 + g_{\mu \nu} N^\mu N^\nu.
\ea
The equations of motions derived from the action (\ref{ADM}) varying
with respect to ${\cal N}$, $N^\mu$ and $g^{\mu \nu}$ yield respectively
\ba
R -2 \L5&=& K^2 - K^\rho_\sigma K^\sigma_\rho \label{EQMON}\\
0 &=& D^\mu K -D^\nu K^\mu_\nu  \label{EQMONN}\\ \label{EQMOG}
G_{\mu \nu} + \L5 g_{\mu \nu}&=& \frac{1}{2}g_{\mu \nu} \left(K^2 - K^\rho_\sigma K^\sigma_\rho\right)  + \frac{D_\mu D_\nu {\cal N} - g_{\mu \nu} D^\rho D_\rho {\cal N}}{\cal N}
- g_{\rho \mu} g_{\sigma \nu} \frac{\partial_y \left\{ \sqrt{-g} \left( K g^{\rho \sigma} - K^{\rho \sigma}\right) \right\}}{{\cal N} \sqrt{-g}} \nonumber \\
&&- \frac{2}{N}\left\{D_\nu \left(N_\mu K\right)
- D_\rho\left(K^\rho_\nu N_\mu\right) - \frac{1}{2} g_{\mu \nu}
D^\rho\left(K N_\rho \right) + \frac{1}{2} D^\rho \left(N_\rho K_{\mu \nu}\right)\right\} \nonumber \\
&&+2 \left( K^\rho_\mu K_{\rho \nu} - K K_{\mu \nu}\right),
\ea
where $G_{\mu \nu}$ is the Einstein tensor for four dimensional metric
$g_{\mu \nu}$, and $K$ is defined by $K \equiv K_{\mu \nu} g^{\mu \nu}$.
When $N_\mu$ vanishes, equations (\ref{EQMON}) and (\ref{EQMOG}) reduce respectively to
\ba
R-2 \L5 &=& {1\over 4
{\cal N}^2}g'_{\mu\nu}g'_{\alpha\beta}(g^{\mu\nu}g^{\alpha\beta}
-g^{\mu\alpha}g^{\nu\beta}), \label{EQN5} \\
G_{\mu\nu}+\L5 g_{\mu\nu}&=&
{1\over {\cal N}}\left(D_\mu D_\nu {\cal N}
-g_{\mu\nu}\  g^{\rho\sigma}
D_{\rho}D_{\sigma}{\cal N}\right)\nonumber\\
&+& {g_{\mu\nu} \over {\cal N}^2} \left(
-{1 \over 8} (g'_{\alpha\beta}g^{\alpha\beta})^2
+{3 \over 8} g'_{\alpha\beta}g'_{\lambda\delta}g^{\delta\beta}
g^{\lambda\alpha}-{1 \over 2} g"_{\alpha\beta}g^{\alpha\beta}
+{1 \over 2} {{\cal N}'\over {\cal N}}g'_{\alpha\beta}g^{\alpha\beta}\right)
\nonumber\\
&+&{1 \over 4{\cal N}^2} g'_{\mu\nu}(g'_{\alpha\beta}g^{\alpha\beta})
+{1 \over 2{\cal N}^2}g"_{\mu\nu}-{{\cal N}' \over 2{\cal N}^3}g'_{\mu\nu}-
{1 \over 2{\cal N}^2}g'_{\mu\alpha}
g'_{\nu\beta}g^{\alpha \beta}, \label{EQg5}
\ea
where $G_{\mu \nu}$ is the Einstein tensor of
the 4D metric $g_{\mu
  \nu}$.
We then consider simple solutions of the 5D Einstein equations,
 which are
slicings of the 5D space-time by 4D Minkowski space-time.
We write accordingly  ${\cal
  N}={\cal N}(y)$,
$g_{\mu\nu}(x^\alpha,y)=\Omega(y) \eta_{\mu \nu}$,
where $\eta_{\mu \nu}$ is a 4D Minkoswski metric,
and we keep $N_\mu=0$.
Then equations (\ref{EQMONN})
are identically satisfied and
the equations (\ref{EQg5}) and (\ref{EQN5})  reduce to
\ba
\label{EQgsimp}
\frac{1}{\N} \left(\frac{\Omega'}{\N }\right)' &=&
-\frac{2}{3}\L5 \Omega,\\
\left({\Omega'\over \ \N}\right)^2&=& -{2\over
3}\L5 \Omega^2 \label{EQNsimp}.
\ea
One notes that the above equations are not independent,
the first being a consequence of the second. This is of
course a simple consequence of the $y$
reparametrization invariance of the 5D Einstein-Hilbert action
(\ref{5DEH}) and tells us that we cannot  solve both for
${\cal N}$ and $\Omega$ out of the equations of motion.
Indeed by choosing a
Gaussian Normal gauge, defined here simply
by setting ${\cal N}$ to 1, one can get the following
solution for the
5D line element
\ba
ds^2 &\equiv& \tilde{g}_{AB}^{(1)} dX^A dX^B  \nonumber \\
&=& dy^2 + \exp \left( \epsilon y \sqrt{-\frac{2}{3} \L5}
\right)
dx^\mu dx^\nu \eta_{\mu \nu}, \label{line1}
\ea
where we have assumed
 $\L5$ to be negative.
These solutions are well known parametrization of a Poincar\'e patch of
$AdS_5$. If one instead puts an absolute value on $y$, in the above
line element, and considers then the metric $\tilde{g}_{AB}^{(2)}$
given by
\ba \label{line2}
ds^2 &\equiv& \tilde{g}_{AB}^{(2)} dX^A dX^B  \nonumber \\
&=& dy^2 + \exp \left( \epsilon |y| \sqrt{-\frac{2}{3} \L5} \right) dx^\mu dx^\nu \eta_{\mu \nu},
\ea
one obtains the well known
Randall-Sundrum \cite{RS} type of metric which are simply given by gluing two
identical patches of the space-time parametrized by metric
$\tilde{g}_{AB}^{(1)}$ along
a brane of positive ($\epsilon = -1$) or negative ($\epsilon = +1$)
tension. This requires a fine tuning between the brane tension
$\tilde{\sigma}$ and the bulk (negative) cosmological constant $\L5$ given by
\ba\label{finetu}
\tilde{\sigma}^ 2 = -6 \Mf^6 \L5,
\ea
where normalization
of the brane tension is set by its action,
\ba \label{BRANEACT}
-\int d^4 x dy \sqrt{-g} 2 \delta(y) \epsilon\tilde{\sigma},
\ea
that was added to
the Einstein-Hilbert action\footnote{Note that we did not include any Gibbons-Hawking term in the brane action, since this term is cancelled by the integration by part to go from (\ref{5DEH}) to (\ref{ADM})} (\ref{ADM}) to get (\ref{line2}).
Another parametrization $\tilde{g}^{(3)}_{AB}$
of the space time described by the line element (\ref{line1})
that we will use in the following, is given by
\ba \label{line3}
ds^2 &\equiv& \tilde{g}^{(3)}_{AB} dX^A dX^B \\
&=& -\frac{3}{2 \tilde \Lambda} \frac{dz^2}{(z+\tilde{z}_0)^2} + Z_0(z+\tilde{z}_0)  \eta_{\mu \nu} dx^\mu dx^\nu,
\ea
where $\tilde{z}_0$ and $Z_0$ are some constants, and
$z$  is related to $y$ of equation (\ref{line1})
 by
$Z_0(z+\tilde{z}_0) = \exp \left(\epsilon y \sqrt{-2/3\tilde \Lambda}\right)$.

\section{Theories with a discretized extra dimension}
We are now turning to  the type of theories which will be of interest in this work. Those can be obtained from the
Einstein-Hilbert action (\ref{ADM}) where one
discretizes the continuuous coordinate $y$ with a spacing $a$ between two adjacent sites (labelled by an index $i$).
It was shown in reference \cite{DM1} how to obtain such a discretization,  maintaining $y$ dependant 4D gauge invariance on each sites. This can be done by
the mean of {\it link fields} $X^\mu(i,i+1;x)$ \cite{Arkani-Hamed:2002sp}, mapping between site $i$ and site $i+1$, which were explicitly build out of the $5D$ metric in reference \cite{DM1}. One can proceed as follows, we first note that the $y$ derivatives arising in action (\ref{ADM}) only appear in the extrinsic curvature $K_{\mu \nu}$ (as defined in Eq. (\ref{DEFK})). The latter can be expressed in term of the Lie derivative ${\cal L}_{D_y}$ along the vector field $D_y$ defined by $D_y = \partial_y - N^\mu \partial_\mu$, such that
\be
K_{\mu \nu} = \frac{1}{2 {\cal N}} {\cal L}_{D_y} g_{\mu \nu}.
\ee
To discretize action (\ref{ADM}) in a way which conserve reparametrization invariances on each sites, we then simply replace every Lie derivative ${\cal L}_{D_y}$, acting on a tensor $T(x^\mu,y)$,
 by its discrete counterpart $\Delta_{\cal L}$ defined as
\be \label{DIFFIN}
\Delta_{\cal L} T_i={ W(i,i+1)T_{i+1}-T_i \over a},
\ee
where the action of the transport operator $W(i,i+1)$ on the
components of tensor $T$ is given by
\be
\left[W(i,i+1)T_{i+1}\right]_{\mu_1,\dots \mu_r}(x)=
\partial_{\mu_1}X^{\nu_1}\dots\partial_{\mu_r}X^{\nu_r}T_{i+1}\left(X^{\mu}(i,i+1;x)\right)_{\nu_1,\dots\nu_r}.
\ee
The transport operators  $W(i,i+1)$ are generated by the shift
vector fields \cite{DM1}
 and at leading order in the transverse lattice
 spacing $a$, $X^\mu$ is given by
\be
X^\mu(i,i+1,x) = x^\mu + a N^\mu(y_i,x) + {\cal O}(a^2),
\ee
where $y_i$ is the $y$ coordinate of site $i$.
If one then  makes a gauge choice such that $X^\mu(i,i+1,x) = x^\mu$, one is lead to consider theories  of a set of 4D metrics $g^i_{\mu \nu}$,
and 4D scalar lapse field $\N_i$
with actions of the form
\ba\label{ACT}
S[g_i,\N_i] = \Sigma_i \MM^2 \int d^4x \sqrt{-g_i} \N_i \left( R(g_i) - 2 \Lambda \right)
- \int d^4x \frac{\MM^2}{\N_i} V(g_i,g_{i+1}),
\ea
where $V(g_i,g_{i+1})$ is an interaction term between the metrics
$g^i_{\mu \nu}$ and
$g^{i+1}_{\mu \nu}$, and $\MM$ is a mass scale which sets the coupling scale between the metric on a given  site and matter sources
that one may wish to put on the same site. If one insists in keeping the link with the 5D theory,
one should  verify the equation of motion for $X^\sigma$. The latter read (in the gauge $X^\mu = x^\mu$)
\ba \label{EQMONDIS}
0&=&\frac{2 {\cal N}_i}{\sqrt{-g_i}}
\partial_\mu\left( \frac{\sqrt{-g_i}}{{\cal N}_i} g_{\sigma \nu}^{i+1}
 \left(g_{\alpha \beta}^{i+1} - g_{\alpha \beta}^i \right)
 \left( g^{\mu \nu}_i g_i^{\alpha \beta} - \nonumber
g^{\mu \alpha}_i g^{\nu \beta}_i \right) \right) \\&&
- \left(\partial_\sigma g_{\mu\nu}^{i+1} \right)
\left( g_{\alpha \beta}^{i+1} - g_{\alpha \beta}^i \right)\left(
 g^{\mu \nu}_i g_i^{\alpha \beta} -
g^{\mu \alpha}_i g^{\nu \beta}_i \right)
\ea
and reduce to equation (\ref{EQMONN}) in the continuum limit.
This equation is somehow similar
to a Kaluza-Klein consistency
condition \cite{KKcons,Duff:1984hn}.
Note that the index $i$ can be envisioned as
labelling {\it theory space} sites in the spirit of the deconstruction
program of Ref. {\cite{Arkani-Hamed:2002sp}, but theories under
  investigation here can also be considered without an explicit
  reference to a continuum limit, simply as
  theories of {\it multigravity}
\cite{Isham:gm,Damour:2002ws}. In the latter case, one does
not have to consider equation
(\ref{EQMONDIS}). We will, for most of the cases discussed
in this paper, not include any matter fields so that each of
the sites $i$
will only be considered endowed with a cosmological
constant $\Lambda$
and will consider various possible interaction terms given by
\ba \label{pot1}
V_1(g_i,g_{i+1})&=& -\frac{m^2}{4}
\sqrt{-g_i}\left( g_{\mu \nu}^{i+1} -
g_{\mu \nu}^i \right)\left( g_{\alpha \beta}^{i+1} -
g_{\alpha \beta}^i \right) \left( g_i^{\mu \nu}
g_i^{\alpha \beta} - g_i^{\mu \alpha} g_i^{\nu \beta} \right),
\\ \label{pot2}
V_2(g_i,g_{i+1})&=&  -\frac{m^2}{4} \sqrt{-g_i}
\left( g^{\mu \nu}_{i+1} - g^{\mu \nu}_i \right)
\left( g^{\alpha \beta}_{i+1} - g^{\alpha \beta}_i \right)
\left( g^i_{\mu \nu} g^i_{\alpha \beta} - g^i_{\mu \alpha}
g^i_{\nu \beta} \right),\\ \label{pot3}
V_3(g_i,g_{i+1})&=& \frac{m^2}{4} \sqrt{-g_i}
\left( g^{\mu \nu}_{i+1} - g^{\mu \nu}_i \right)
\left( g_{\alpha \beta}^{i+1} - g_{\alpha \beta}^i \right)
\left( g^i_{\mu \nu} g_i^{\alpha \beta} -
\delta^\mu_\alpha \delta^\nu_\beta \right),
\ea
where $g_i^{\mu \nu}$ is the inverse metric of $g_{\mu \nu}^i$.
With such choices of interaction terms, action (\ref{ACT}) is a
simple minded discretization of the 5D pure gravity
Einstein-Hilbert
action (\ref{5DEH}),
where one has set $N_\mu$ to zero.
This can be seen
explicitly  using the following identification
\ba
 m^2  &=& \frac{1}{\a^2} \label{mD},\\
\MM^2 &=& \Mf^3 \a, \\
\Lambda  &=& \L5 \label{LD}, \\ \label{gD}
g_{\mu \nu}^i(x^\mu) &=& \tilde{g}_{\mu \nu}(x^\mu,y_i),\\
\N_i(x^\mu) &=& \N(x^\mu,y_i) \label{Ni},
\ea
where $\a$ is the size of the discretization step along $y$, and
$y_i = i \a$.
Note that each of the three potential
(\ref{pot1}), (\ref{pot2}) and (\ref{pot3})
corresponds to a different way to implement
the discretization
procedure\footnote{For simplicity,
we have written here only the equation
of motion (\ref{EQMONDIS}) corresponding to the potential $V_1$}
 outlined in the beginning of this section,
 depending on whether one discretizes
 $K_{\mu \nu} = {\cal L}_{D_y} g_{\mu \nu} / 2 {\cal N}$ or
  $K^{\mu \nu} = - {\cal L}_{D_y} g^{\mu \nu} / 2 {\cal N}$.

\subsection{No lapse field and brane space-time with no brane source} \label{NOBRA}
We first set from the beginning the {\it lapse}  fields $\N_i$ to one
in the multigravity action  $S[g_i,\N_i]$ and consider solutions to the
equations of motion derived from the simplified action
for the metrics
$g^i_{\mu \nu}$, $S[g_i,1]$. We wish here to compare these
solutions
with solutions of the continuum theory defined by action
(\ref{ADM})
and seek  solutions of the form
\ba\label{ansatz}
g_{\mu \nu}^i = \Omega^i \eta_{\mu \nu},
\ea
with $\Omega^i$ constants.
Let us first consider the case of the
interaction term $V_1$. After a straightforward calculation,
the equation of motion for the metric $g_i$ reduces to
\ba \label{SEQ}
-\lambda = -2 + f_i +  \left(f_{i-1}\right)^{-1},
\ea
with $\lambda = {2 \Lambda / 3 m^2}$, and
$f_i = \Omega_{i+1} / \Omega_i$.
Equation (\ref{SEQ}) defines a sequence
(see figure \ref{fig1}) which has two fixed
points
(for $\lambda$ obeying $\lambda (-1+ \lambda/4 ) >0)$
$f_+$ and $f_-$ given by
\ba
f_{\pm} = \left(1-{\lambda \over 2} \right)
\pm \sqrt{\lambda \left({\lambda \over 4} -1 \right)}.
\ea
These two fixed points are each located on a different side
from 1 (for negative $\lambda$) , which is the double
root of the
fixed point equation for $\lambda = 0$.
Let us first consider the solutions ${\cal F}_\pm$ defined by
$\forall i, f_i = f_{\pm}$. This translates into
\ba
\Omega^j_\pm = \Omega^0_\pm \left( f_\pm \right)^{y_j /\a}.\label{sol2}
\ea
We then look at the limit of this solution when the discretization
step $\a$ goes to zero.
>From equations (\ref{mD}) and (\ref{LD}), one has
$\lambda = \frac{2}{3} \L5 \a^2,$ which gives in the
limit of small $\a$
\ba
f_\pm \sim 1 \pm \sqrt{-\lambda} \sim  1 \pm \sqrt{-\frac{2}{3}
  \tilde{\Lambda}} \a, \label{FPOINT},
\ea
leading to,
\ba
g_{\mu \nu}^{j} \sim \Omega^0_\pm
 \exp \left( \pm \sqrt{-\frac{2}{3} \tilde{\Lambda}} y_j \right)
 \eta_{\mu \nu}.
\ea
This  matches  the metric (\ref{line1}), and
the fixed point solutions
 ${\cal F}_\pm$, in the limit of
 small $\a$ thus each corresponds to  a discretized version of a
Poincar\'e patch of $AdS_5$.

However equation (\ref{SEQ}) has  more interesting solutions, it can
indeed  be  exactly solved to yield
\ba \label{EXACT1}
\Omega^j =
\left( (f_+)^j + {\cal K} (f_-)^j\right) \frac{\Omega^0}{1+ {\cal K}},
\ea
where ${\cal K}$ is an integration constant, related to $f_0$ by
\ba
{\cal K} = -\frac{f_+- f_0}{f_--f_0}.
\ea
Let us consider the case where ${\cal K}$ is  positive,
 this is true if we
choose $f_0$ to lie in between $f_-$ and $f_+$.
Note first that, for a fixed value of $\a$ different from zero,
one can choose any positive value of ${\cal K}$. One easily finds that the limiting behaviour of the metric on site
$i$, when $\a$ is send to zero
is given by
\ba \label{NEGTEN}
g_{\mu \nu}^i \sim  \frac{2 \Omega^0 \sqrt{\cal K}}{(1+{\cal K})}
{\rm cosh} \left(\sqrt{-\frac{2}{3} \Lambda} (y_i - \Upsilon)\right)
\eta_{\mu \nu},
\ea
 where $\Upsilon$ is a constant given by
\ba
\Upsilon = \sqrt{-{3 \over 8 \Lambda}} {\rm ln}({\cal K}).
\ea
It turns out that the solution (\ref{NEGTEN}) has  exactly
the same asymptotics\footnote{Note however, that the exact expression
  (\ref{EXACT1}) does not approach the asymptotic form given by
(\ref{NEGTEN}) uniformly with $i$, i.e., for a fixed value of
  $\a$, one can make the difference between $a_i$ as given by
 (\ref{EXACT1}) and the asymptotic form read from
(\ref{NEGTEN}) as large as one wishes going to large value of
  $y_i$. On a given compact interval though, say of the form
  $[-y_k,y_k]$ ($k$ being fixed), the convergence is uniform.}
as $y_i$ goes to $\pm \infty$
as the Randall-Sundrum metric (\ref{line2}) with
a negative tension brane placed at $y=\Upsilon$.
This is a quite remarkable feature since no brane has been
considered. In particular no fine tuning of the form
(\ref{finetu}) is necessary.

One might be concerned about knowing to what extent
this result depends on the form of the discretization scheme.
We now investigate this question by searching for
solutions of the form (\ref{ansatz}) with $V$ taken of the form
\be
V = {\cal C}_1 V_1 + {\cal C}_2 V_2 + {\cal C}_3 V_3,
\ee
by changing $m^2$ we can always assume that
${\cal C}_1 + {\cal C}_2 +
{\cal C}_3= 1$. This choice also insures that the
multigravity action $S[g_i,1]$ has the action (\ref{ADM})
for a limit when
$\a$ goes to zero, with the same identification as in equations
(\ref{mD}-\ref{Ni}). It is then straightforward to
work out the equations of motion for the metrics
$g_{\mu \nu}^i$ and get for the ansatz (\ref{ansatz})
the following expression replacing
equation (\ref{SEQ})
\be \label{SEQBIS}
-\lambda = G_1(f_i) + G_2(f_{i-1}),
\ee
with
\ba
G_1(u) &=&-4\aa-3\b +2 + u
 \left( -\frac{1}{2} + \frac{3}{2}\aa +
\frac{1}{2}\b \right) \nonumber \\
&&+ \frac{1}{u}
\left( -\frac{3}{2} + \frac{3}{2} \aa + \frac{9}{2} \b \right)
 -\frac{2}{u^2} \b,  \\
G_2(u) &=& \frac{1}{u}
\left( -\frac{1}{2} +\frac{3}{2}\aa + \frac{1}{2}\b \right)
+ \frac{1}{u^3}
\left( \frac{1}{2}  -\frac{1}{2}\aa - \frac{3}{2}\b \right)
+ \frac{\b}{u^4}.
\ea
It is easy to see that $u=1$ is still a double root to
the equation
$G_1(u) + G_2(u)=0$. For $\lambda < 0$, this double root
degenerates into two root $f_{\pm}$, which are given
(for small enough $\lambda$) by
\be
f_\pm -1 \equiv \delta_{\pm} \sim \pm \sqrt{-\lambda}.
\ee
Notice that the above expression is independent of $\aa$ and $\b$, as could have been expected, and matches the one found in equation (\ref{FPOINT}).
While in the vicinity of these roots the sequence $\delta_i$
defined by
$f_i = 1 + \delta_i$ verifies
\be
\delta_i = \delta_{i-1} + \delta_\pm^2 - \delta_{i-1}^2.
\ee
This leads to the same kind of solution as above:
namely a solution
with the same asymptotics as  the Randall-Sundrum solution with a
negative tension brane.

There is an easy way to understand these solutions by
comparing the
equations of motion of the continuum theory
(\ref{EQgsimp}-\ref{EQNsimp}) to the ones of the
multigravity theory
(\ref{SEQ}) (or (\ref{SEQBIS})). Indeed, in the limit where $\a$
goes to zero, equation (\ref{SEQ}) (or (\ref{SEQBIS}))
reduces to
equation (\ref{EQgsimp}). If one then looks for solutions of
(\ref{EQgsimp}), with $\N$ set to one, one finds that
the most general
solution is given by a linear combination of
$\exp \left(k y \right)$ and $\exp (-ky)$ with $k$ given by
$\sqrt{- 2 \Lambda /3}$. This matches what is found
in the discretized
theory. However, in the continuum theory,
the equation of motion for $\N$, namely
 equation (\ref{EQNsimp}),
 allows to keep the decreasing or increasing exponential,
but not a combination of the two. This also enables to
understand  that we did not find the solution
with a positive tension brane (which would  have been very
interesting in many respects). Indeed, a linear combination of
exponentials $\exp \pm ky$ (with positive coefficients) is
an increasing function for large positive $y$ and a decreasing
function for large negative $y$. So that the asymptotic {\it jump} of
the first $y$-derivative of the 4D metric across the {\it brane},
defined as $[\Omega'(+ \infty)-
\Omega'(-\infty)]/[\Omega(+ \infty)+
\Omega(-\infty)]$, is necessarily positive.
This is turn  means that the brane
tension has to be negative, as can be seen
 from the
junction conditions.
In the next subsection, we show how to recover the
positive tension
brane solution, introducing a brane source with fine-tuned
tension. We
then, in the last subsection, turn to investigate the
discretized theory
where the lapse field $\N$ is kept, which seems to be required if one
wants to approach closer solutions to the continuum theory.
Let us
underline here, however, that the solution (\ref{EXACT1}) is a
perfectly honest  solution of the multigravity theory defined
by action $S[g_i,1]$.
To complete this discussion, let us add that the solution
(\ref{EXACT1}), and more generally solution of equation
(\ref{SEQBIS}), can also differ much more dramatically
from solutions
of the 5D continuum theory. In particular, the fixed
points solutions
${\cal F}_\pm$ are very tuned solutions. Indeed any departure
of $f_0$, say as an ``initial'' choice
\footnote{The term ``initial''
  might be somehow misleading here, since the sequence $f_i$
  is also
continued in the ``past'', $i \rightarrow -\infty$, from the
``initial'' value $f_0$.},
 from $f_\pm$ leads
 either to the negative brane-like solution, or to other type
 of solution where the signature of the
4D metric can change (as can be seen from equation (\ref{EXACT1}) with ${\cal K} <0$).

\subsection{Positive tension brane and fine tuning} \label{BRA}
In this subsection (and only here), we include
a brane source\footnote{Namely, in contrast to the previous cases,
we are now explicitely breaking the $i$ translation invariance by
allowing the "cosmological constant" (or tension)
of a given site (or brane) to differ from the one of its
neighbours.}, located at fixed $y$,  in the discretized theory
considered
above. We wish to recover the discretized Randall-Sundrum
metric with a positive tension brane, and discuss how the
fine tuning
condition (\ref{finetu}) is translated in the discretized theory. To do so, we simply add to the action
 $S[g_i,1]$, considered above with the potential $V_1$,
 the discretized version of the action
 (\ref{BRANEACT}) which can be simply taken to be
\ba
-\int d^4 x \sqrt{-g_0} {2 \sigma },
\ea
with $\sigma = \tilde{\sigma}$, and we have chosen the
brane to be
 located on site $i=0$.
The only difference with the analysis done in the
previous subsection
 is that the cosmological constant $\Lambda$ of site $i=0$
 is now
 replaced by $\Lambda + \sigma/\MM^2$. We now seek to
reproduce    the Randall-Sundrum metric (\ref{line2})
 with $\epsilon = -1$ in the continuum limit. For $i \neq 0$,
 the equations of motion of
 $g^i_{\mu \nu}$ lead to the same equation as (\ref{SEQ}) that we
 rewrite
\ba
f_i = F_\lambda(f_{i-1}),
\ea
with $F_\lambda(f_{i-1}) \equiv 2 - \lambda - (f_{i-1})^{-1}$.
If we consider first the sequence of $f_i$ with $i \geq 0$,
then we
 need this sequence to lead to a decreasing $a^i$,
 for large $i$,
 however the general solution (\ref{EXACT1}) shows this is never the
 case unless we choose $f_0 = f_-$. Similarly, considering now the
 sequence of $f_i$, with $i\leq -1$,  we need to choose $f_{-1}$
 to be equal to $f_+$ in order to get an  increasing function of $y$, as
 $y$ becomes large and negative. In order to reproduce
 the Randall-Sundrum solution, with a positive tension brane, one thus
 needs to jump from $f_{-1} = f_+$ to $f_0 = f_-$. However $f_{-1}$
 and $f_0$ are related by the $g_{\mu \nu}^0$ equation of motion
leading to
\ba
f_0 = F_{\lambda + 2\sigma/ 3 m^2 \MM^2}(f_{-1}).
\ea
Demanding then $f_- =  F_{\lambda +2 \sigma/ 3 m^2 \MM^2} (f_+)$ leads to a
 fine tuned value for $\sigma$ given by (see figure \ref{fig1})
\ba
\sigma =  3m^2 \MM^2 \sqrt{\lambda\left(\frac{\lambda}{4} -1 \right)}.
\ea
This in turn, leads back in the small $\a$ limit to the constraint (\ref{finetu}).

\begin{figure}
\psfrag{a}{{\footnotesize $f_i$}}
\psfrag{b}{{\footnotesize $f_{i+1}$}}
\psfrag{c}{{\footnotesize $f_{i+2}$}}
\psfrag{g}{{\footnotesize $f_+$}}
\psfrag{f}{{\footnotesize $f_-$}}
\psfrag{l}{{\footnotesize $-\lambda$}}
\psfrag{m}{{\footnotesize $2\sigma / 3 m^2 M_P^2 $}}
\psfrag{x}{{\footnotesize $u$}}
\psfrag{y}{{\footnotesize $v$}}
\epsfig{file=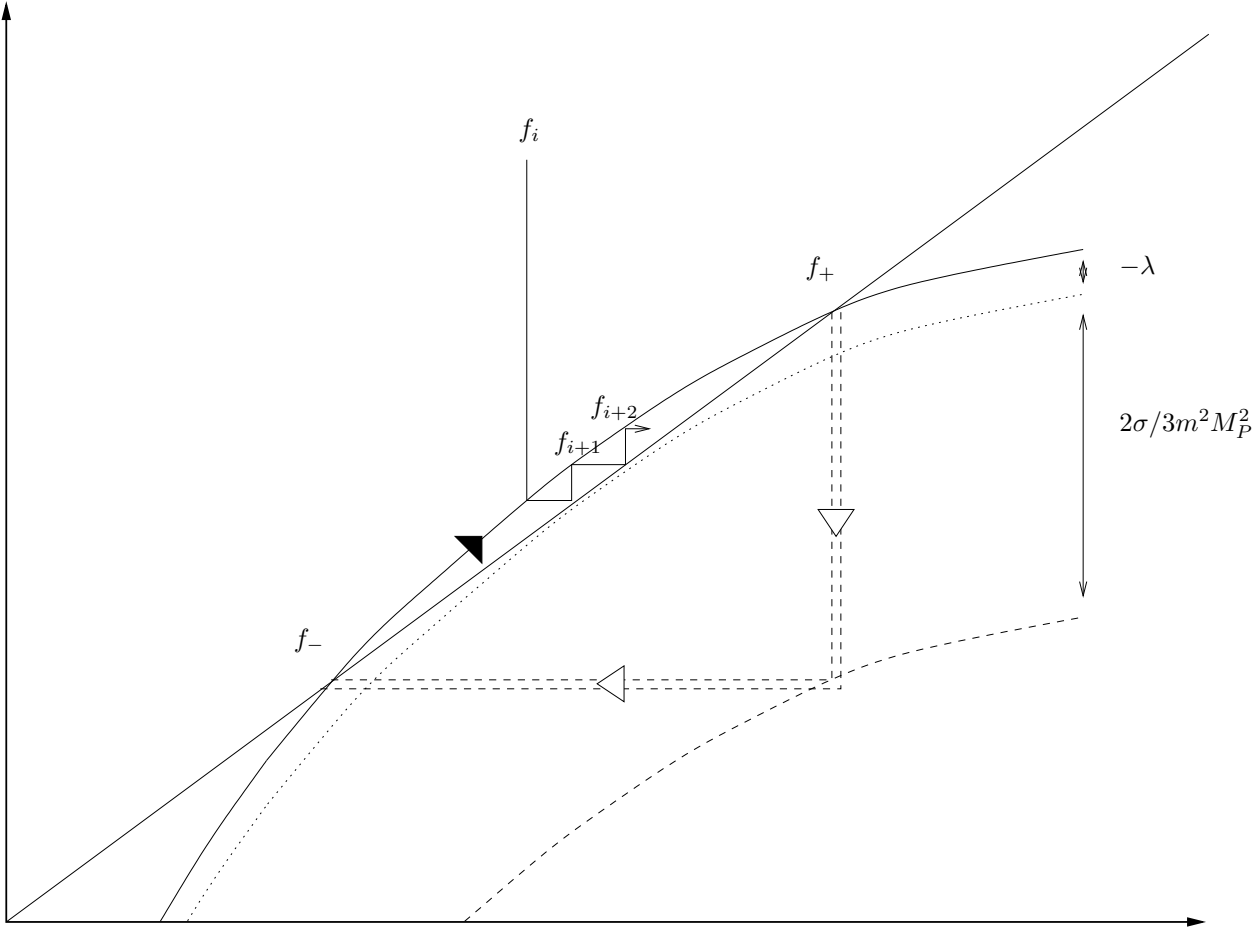, width=1.0\textwidth}
\caption{Depiction of the resolution of equation (\ref{SEQ}) without
  (subsection (\ref{NOBRA})) and with (subsection (\ref{BRA})) a brane
  source included. Equation (\ref{SEQ}) defines a sequence which for
  $\lambda < 0$ and small, is characterized by the upper solid
  curve. The latter curve is
  representing the function $F_\lambda(u)$ which defines the sequence by
  the recursion relation $f_i = F_\lambda(f_{i-1})$. The solid oblique
  line is defined by the function $u=v$. One sees that, when no brane
  is included, one can only flow (black arrow) from the fixed point
  $f_-$ to the
  fixed point $f_+$, for an ``initial'' value $f_0$ which lies
  in between the fixed points. This corresponds to the solution
  (\ref{EXACT1}) and to the negative tension brane discretized bulk. When a brane
  source is included, one can go ``backward'' (white arrows and double dashed
  lines) from the fixed point $f_+$ to the fixed point $f_-$
at a given value of the index $i$ corresponding to the brane. For this to happen, the brane tension has to be tuned in
  order for the curve representing $F_{\lambda +2 \sigma/3 m^2 \MM^2}$, the
  dashed bottom curve, to go through the point $(u= f_+, v=f_-$). The
  solution is then the discretized Randall-Sundrum metric with a
  positive tension brane. The dotted curve represents the function
  $F_0$. It is tangent to the $u=v$ line at the point $(1,1)$. It also
  corresponds to the sequence which solution is obtained in equation
  (\ref{SOL1}) of subsection \ref{REPA}.}
\label{fig1}
\end{figure}

\subsection{Lapse field and reparametrization invariance} \label{REPA}
We then turn to another possible discretization of the action
(\ref{ADM}) keeping the lapse field ${\cal N}$ on all sites.
Starting with $V = V_1$, the equations of motion now read
\ba
\sqrt{-\lambda} {\cal N}_i &=& \epsilon_i (f_i-1), \label{EQ1}\\
-\lambda {\cal N}_i &=& \frac{1}{{\cal N}_i}\left( f_i-1 \right) - \frac{1}{{\cal N}_{i-1}} \left( 1 - \frac{1}{f_{i-1}} \right), \label{EQ2}
\ea
where $\epsilon_i = \pm 1$. Fixing $\epsilon_i $ independently of $i$, and replacing ${\cal N}_i$ by its expression as a function of $f_i$ given by equation (\ref{EQ1}) into equation (\ref{EQ2}), one finds that $f_i$ verifies equation
(\ref{SEQ}) with $\lambda =0$, namely
\be
-2 + f_i + \left(f_{i-1}\right)^{-1} =0.
\ee
This can be straightforwardly solved to yield
\ba
g_{\mu \nu}^i&=& I_0(i + i_0) \eta_{\mu \nu} , \label{SOL1}\\
{\cal N}_i^2 &=& \frac{1}{-\lambda (i+i_0)}, \label{SOL2}
\ea
with
$i_0$ and $I_0$ some constants. This exactly reproduces the metric $\tilde{g}_{AB}^{(3)}$, of equation (\ref{line3}) sampled at
successive sites $i$ parametrized by their coordinate $z_i \equiv \a i$ along the $z$ direction. Indeed $g_{\mu \nu}^i$ and ${\cal N}_i$ as defined by equations
 (\ref{SOL1}) and (\ref{SOL2}) are given by $g_{\mu \nu}^i =
\tilde{g}^{(3)}_{\mu \nu} (z_i, x^\mu)$ and ${\cal N}_i^2 =  \tilde{g}^{(3)}_{yy} (z_i, x^\mu)$
with the identifications
$\tilde{z}_0 = i_0 \a$ and $Z_0 = I_0 /\a$. This means in
particular that our discretization scheme reproduces,
 when the discretization step is send to zero,
 solutions of the 5D continuum theory in a given gauge. This is of
 course not a surprise, since our discretized action $S[g_i,\N_i]$
 explicitly breaks reparametrization invariance along the $y$
 direction. It can also be seen by noticing that, in contrast to the
 continuum case, the equation of motion of the metric (\ref{EQ2})
cannot be obtained from the equation of motion for the lapse field
(\ref{EQ1}). This very fact was indeed used to solve for $\N_i$ and
$g_{\mu \nu}^i$. However, note that both equations reduce to equations
(\ref{EQgsimp}) and (\ref{EQNsimp}) in the leading order in
$\a$. So that once the solution
for $\N_i$ (respectively $g_{\mu \nu}^i$) of equations (\ref{EQ1}-\ref{EQ2}) is known, which can be
considered from the point of view of the continuum theory as setting
the gauge, as far as $y$ reparametrization is concerned,
then the solution for the $g_{\mu \nu}^i$ (respectively $\N_i$) agrees with the
solution of the continuum in the small $\a$ limit.

Another property of the solution (\ref{SOL1},\ref{SOL2})
is that the parameter $i_0$ allows one to  avoid the
potential singularity which exists in the continuum.
In fact if $i_0$ is not an integer than the scale factor $a_i$
does
not vanish for all $i$. This is a generic feature of discretized
models.

One can ask whether one can restore the arbitrariness in $\N_i$ by
changing the discretization scheme. It turns out indeed to be
possible choosing the interaction potential $V$ of $S[g_i, \N_i]$ to
depend on metrics on three adjacent sites, and reading
\be \label{POT}
V(g_{i-1},g_i,g_{i+1}) =-
{ m^2 \over 16}\sqrt{-g_i}\left( g_{\mu \nu}^{i+1} -
g_{\mu \nu}^{i-1} \right)
\left( g_{\alpha \beta}^{i+1} - g_{\alpha \beta}^{i-1} \right)
\left( g_i^{\mu \nu} g_i^{\alpha \beta} - g_i^{\mu \alpha}
g_i^{\nu \beta} \right).
\ee
With this interaction term, the action $S[g_i,\N_i]$ also
agrees with
 action (\ref{ADM}) in the continuum limit.
The equations for $\Omega_i$ and $\N_i$ now read
\ba
\Lambda \N_i &=& \frac{3m^2}{8} \left( {\Omega_i-
\Omega_{i+2} \over \Omega_i \N_{i+1} }
+{ \Omega_i-\Omega_{i-2}  \over  \Omega_i \N_{i-1} }
\right) \label{di1}\\
\sqrt{{-\frac{8}{3} \Lambda}} \N_i&=& \epsilon_i m
\left({\Omega_{i+1}-\Omega_{i-1} \over
\Omega_i}\right). \label{di2}
\ea
If one chooses the same value of $\epsilon_i$ for all $i$, then
the first equation is a consequence of the second.
It is thus possible  to ``choose a gauge" by fixing ${\cal N}_i$
and then to find $\Omega_i$ by solving (\ref{di2}),
the resulting
solution will automatically satisfy (\ref{di1}). Let us choose,
for instance, ${\cal N}_i=1=\epsilon_i$, then (\ref{di2}) can be
readily solved
to give
 a solution depending on two parameters
 \be
 \Omega_i=\alpha x_+^i+\beta x_-^i,
 \ee
 where
 \be
 x_{\pm}=\sqrt{-\lambda}\pm\sqrt{1-\lambda}.
 \ee
 Notice that $x_-$ is negative so that the solution with
 $\alpha=0$ has $\Omega^i$ with alternating signs: neighboring
 spacetimes $i$ and $i+1$ have opposite signatures.
 This behavior has no continuum counterpart where a
 singularity is
 reached before the signature change takes place.
When $\beta=0$ the solution coincides, to the first order in
$\a$ with the solution we found previousely in (\ref{sol2})
and so it tends to the $AdS_5$ metric in the continuum.
Notice that had we started to solve (\ref{di1}) with $\N_i=1$
than we would have obtained the analog of the more general
solution (\ref{EXACT1}) but equation (\ref{di2}) would not have been
satisfied.

To summarize, the inclusion of $\N_i$ allows to
recover just the continuum solutions
if we do not insist on having an $\N_i$ undetermined by the
equations of motion\footnote{and we
do only consider solutions which are continuous in the continuum limit};
 and if we do insist, we get extra solutions with no
continuum counterpart.

\section{Effective 4D gravity}
We turn now to determine the gravitational potential
 without the lapse field, and
 consider the discretisation
scheme
given in the previous subsection. We assume here that the
cosmological constant vanishes and so we can perturb around flat
space-time.
To quadratic order in the
metric perturbation $h_{\mu \nu}^i$, the potential
(\ref{POT}) reads
\be \label{QUAD}
V(h_{i-1},h_i,h_{i+1}) =-
{ m^2 \over 16}\left( h_{\mu \nu}^{i+1} - h_{\mu \nu}^{i-1} \right)
\left( h_{\alpha \beta}^{i+1} - h_{\alpha \beta}^{i-1} \right)
\left( \eta^{\mu \nu} \eta^{\alpha \beta} - \eta^{\mu \alpha}
\eta^{\nu \beta} \right).
\ee
In order to diagonalize this interaction, we define
$\check{h}_{\mu \nu}$ by
\ba
\check{h}_{\mu \nu}\left(x^\alpha,\theta \right) =
\sum_{n} h^n_{\mu \nu}(x^\alpha)e^{in\theta}.
\ea
The quadratic action is then an integral over $\theta$
of Pauli-Fierz actions with a continuous mass spectrum given by
\be
m^2(\theta)={\sin^2\theta \over \a^2}.
\ee
And the gravitational potential
 $\Psi_j(r)$,  between two
unit masses  separated by $r$ and placed at sites $i$ and $i+j$,
can be obtained
summing over Pauli-Fierz propagators.
The outcome of the calculation can also be simply understood
from discretizing
the Laplacian equation with the same discretization scheme as
above.
The discretized equation reads
\be
\partial^2 \Psi_j+{{(\Psi_{j+2}+\Psi_{j-2}-2\Psi_j)}\over 4\a^2}
=4\pi G_N\delta(\bf r)\delta_{j,0},\label{dis1}
\ee
where we have reintroduced the Newton constant $G_N$.
The Fourier transform $\check{\Psi}$, as defined above, verifies
\be
\partial^2 \check{\Psi}(r,\theta)-m^2(\theta)\check{\Psi}
(r,\theta)=4\pi G_{N}
\delta(\bf r).
\ee
Notice that the mass spectrum is bounded from above by
the inverse
lattice spacing $\a^{-1}$. A continuous mass spectrum is
reminiscent of the infinite dimensional models of
{\cite{Gregory:2000jc,Dvali:2000rv}. The gravitational
potential can now be
readily put in the form
\be
\Psi_j(r)=-{G_N \over r}
\int_{0}^{\pi}{d\theta \over {\pi}}
e^{-{r \over \a}\sin\theta}
\cos j\theta.
\ee
When $r<<\a$ then the integral can be approximated by
$\delta_{j,0}$ and the potential reduces to the 4D Newtonian
potential
\be
\Psi_j(r)=-{G_N \over r}\left(\delta_{j,0}+
O({r \over \a})\right),
\ \ \ r<<\a.
\ee
When $r>>\a$, then the integral
can be approached by
\be \nonumber
\int_{0}^{\pi/2}{d\theta \over {\pi}}
e^{-{r \over \a}\theta}
\cos j\theta+\int_{\pi/2}^{\pi}{d\theta \over {\pi}}
e^{-{r \over \a}(\pi-\theta)}
\cos j\theta={1\over \pi\a}{r \over
{r^2+\a^2j^2}}(1+(-1)^j+O(e^{-r\over \a})),
\ee
so that the gravitational potential is of the form of a 5D
potential
\be
\Psi_j(r)=-{G_5 \over r^2+(j\a)^2},
\ee
if $j$ is even,
with the 5D gravitational constant being given by
\be
G_5={2G_N \over{\pi \a}}.
\ee
When $j$ is odd the gravitational potential is exponentially
small.
Gravity is thus four dimensional at small length scales and
five dimensional at large  scales. Choosing $\a$ very large
(of the order of the Hubble scale) allows a very simple
modification of gravity at large scales in the spirit of
the models of references \cite{{Kogan:1999wc},{Gregory:2000jc},
{Dvali:2000rv},{Kogan:2000cv}}.
Note that we could have started from a finite
number $N$ of sites. This corresponds to an extra dimension which
is compact with a length scale $R=N\a$.
The graviton spectrum
would have been discrete with a spacing of order $1/N$
and would have remained bounded. In this case,
the four dimensional regime is obtained both for small scales
$r<<\a$ and large scales $r>>R$, the intermediate scales
being five dimensional $\a<<r<<R$.
The qualitative features of
this potential are not sensitive
to the particular form of the discretization scheme we have used.
On the other hand, it is well known
that the tensorial structure of the propagator
of massive spin 2 fields differs dramatically from
the massless one. This leads to the vDVZ discontinuity
at the linearized level which is manifested by,
e.g.,
an order one difference in light
bending.
In this respect, when $R$ is infinite we
expect the discontinuity to be present since
we have a continuous spectrum of massive gravitons.
This is similar to the brane models with an extra infinite
dimension \cite{{Kogan:1999wc},{Gregory:2000jc},
{Dvali:2000rv},{Kogan:2000cv}}. When $R$ is finite,
however, the spectrum is discrete and there is no discontinuity.

\section{Cosmological solutions}
The discussion of preceeding subsections enables  to find
easily some cosmological solutions for the type of models
considered here. It is the purpose of this last section to
discuss those solutions. We look for solutions where the line
element of site $i$ takes the form \be
ds^2_i=-B_i^2(t)dt^2+A_i^2(t)d\vec{x}^2. \ee The equations
of motion
derived from action (\ref{ACT}) (where we consider ${\cal
N}_i$ as a non dynamical field set to $1$) with the potential
$V_1$ read \ba 3{\dot A_i^2\over A_i^2}-\tilde
\Lambda B_i^2
&=&
-\frac{3 B_i^2}{2 \a^2} \left\{1 + \frac{B^2_{i+1}}{2 B^2_i} -
 \frac{A^2_{i+1}B^2_{i+1}}{2 A^2_i B^2_i}\nonumber-
\frac{3 A^2_{i+1}}{2 A^2_i} +  \frac{A_{i+1}^4}{2 A^4_i}
\right\} \\
&&-\frac{3 B_i^2}{2 \a^2} \left\{ \frac{B_i}{A_i}
\frac{A_{i-1}}{B_{i-1}}
\left(\frac{A^2_i-A^2_{i-1}}{A^2_i} \right) \right\}
 + M_{(4)}^{-2} B^2_i \rho_i, \\
 -{\dot A_i^2\over A_i^2}-2{\ddot A_i\over A_i}+2 \nonumber
{\dot A_i\dot B_i\over A_i B_i}+\tilde \Lambda B_i^2&=&
-\frac{ B_i^2}{2 a^2 A_i^2}
\left\{
-\frac{B^2_{i+1} A^2_{i+1}}{2 B_i^2} +  \frac{A_{i+1}^4}{2 A^2_i}
+ \frac{3}{2} A^2_{i+1} - 3 A_i^2 +
\frac{3 B_{i+1}^2 A^2_i}{2 B^2_i}\right. \\
&& \left.
 - A_i^2 \frac{A_{i-1}B_{i-1}}{A_i B_i}\left(\frac{B^2_i}
 {B^2_{i-1}} + 2
 \frac{A^2_i}{A^2_{i-1}}-3 \right)\right\}+M_{(4)}^{-2}
 p_i B_i^2,
\ea
where we have included the possibility for each site $i$ to
host matter with energy density $\rho_i$ and pressure $p_i$
(that is to say, matter, in the form of a perfect fluid,  is assumed to be confined on each site).
We then look for factorized solutions of the form
\ba
A^2_i(t)&=&\alpha_i A^2(t), \nonumber\\
B_i^2(t)&=&\beta_i B^2(t), \label{ANSATZ} \ea where $\alpha_i$ and
$\beta_i$ are time-idependent, we also set $B$ to one by a suitable
time redefinition.
We then further simplify the problem by letting $\alpha_i =
\beta_i$, and look for  solution where the metric on each site
is de Sitter, and the sites are only endowed with a cosmological constant $\sigma$. This means  $\rho_i$ obeyes $\rho_i = - p_i=\sigma$ and does not depend on $i$. In this case,
the equations of motion reduce to \ba
\frac{\dot{A}^2}{A^2} &=& H^2,  \nonumber\\
\frac{3}{2a^2} \left(\alpha_{i+1}+ \alpha_{i-1}-2 \alpha_i \right)
&=& 3 H^2 - \left(\tilde{\Lambda}+ M_{(4)}^{-2} \sigma
\right).\label{EQCOSMOBIS} \ea These equations can easily be
solved. Let us first discuss the case where $(\tilde{\Lambda} +M_{(4)}^{-2} \sigma)$ does not vanish. In this case, defining $\gamma$ by
$\gamma = 3 H^2 /\left( \tilde{\Lambda} + M_{(4)}^{-2} \sigma \right)$,
the sequence $f_i  \equiv (\alpha_{i+1} -
\gamma)/(\alpha_{i}-\gamma)$ obeyes equation (\ref{SEQ}) with
$\lambda$ given by $\lambda \equiv 2( \tilde{\Lambda} +  M_{(4)}^{-2}
\sigma)a^2/3$. So that the solutions are readily obtained from
those of section 3. In the case where $(\tilde{\Lambda} + M_{(4)}^{-2}
\sigma)$ vanishes, on the other hand, the solution to equation
(\ref{EQCOSMOBIS}) is given by \ba\alpha_i = \alpha_0 +
\label{SOLCOS}
\tilde{\alpha}_0 i + H^2 a^2 i^2,\ea where $\alpha_0$ and
$\tilde{\alpha}_0$ are some arbitrary constants.
Interestingly, this solution has de Sitter 4D space-time on
each sites, but vanishing cosmological constants.
This is in analogy with what is happening in the brane-induced
gravity model
\cite{Dvali:2000hr,ACCDGP} as is the behavior of the
Newtonian gravitationnal potential that was discussed
in the preceeding section. However in contrast to
brane-induced gravity, the curvature of de Sitter
space-time is here arbitrary.
Note that, had we kept the $X^\mu_i$ fields as a dynamical
variables,
one should  have also considered their equations of
motion. The latter, for the ansatz (\ref{ANSATZ}), reduce to
\be { \dot A \over
A}\left(2{ \alpha_{i+1} \over \alpha_i}-3\right) \left({
\alpha_{i+1} \over \alpha_i}-{ \beta_{i+1} \over \beta_i}\right)
=0
.\label{xx} \ee
This equation is satisfied for our choice
$\alpha_i = \beta_i$. However, in a similar way as was discussed
previously,
solution (\ref{SOLCOS}) does not verify (for $H \neq 0$)
the equations of motion for ${\cal N}_i$ which reads
(for ${\cal N}_i =1$)
\ba
\frac{6}{\alpha_i}\left(\frac{\dot{A}^2}{A^2}+
\frac{\ddot{A}}{A}\right) = \frac{3}{a^2}\left
( \frac{\alpha_{i+1}-\alpha_i}{\alpha_i}\right)^2.
\ea
Indeed in the continuum theory, there is no solution
equivalent to
 (\ref{SOLCOS}) with arbitrary $\alpha_0$ and
 $\tilde{\alpha}_0$; rather one finds a related solution
 describing  a slicing
of 5D Minkowski space-time by  4D de Sitter space-times
which reads
\ba
ds^2 = (1 + H y)^2\left( -dt^2 + e^{2Ht} d\vec{x}\right) + dy^2.
\ea
This solution is conveniently used to described
the space-time on one side of a domain-wall,
which entire space time is given by the above
solution where $y$ is replaced by $|y|$ \cite{DW}.

\section{Conclusions}
In this paper, we have investigated some exact solutions of 5D General
Relativity with a discrete fifth dimension,
and compared them
to solutions of the continuum theory. Depending
on the discretization
scheme  used, we have shown that some of the solutions of the
discrete theory exactly match those of the continuum,
while others do
not.
In general, the discretization
 explicitly breaks reparametrization along the discrete
dimension.
This gets reflected in the fact that for solution we have
considered, corresponding to slicing of $AdS_5$ by Minkowski space,
the equations of motion of the discrete theory enable
in general to
determine both the lapse field and the 4D metric.
This is to be contrasted
with the continuum theory, where the lapse field cannot
 be determined
by the equations of motion.
However, we have
shown that one can find a given discretization scheme
in which it
remains undetermined.
More importantly, some of the solutions of the discrete
theory exibit very dramatic differences with those of the
continuum. As shown here, one can find signature change
of the 4D
metric but also avoidance of singularities
which would be present in the continuum. In addition we
 have also
found a brane world looking bulk space time, with no brane
source in
the equation of motion and no fine tuning. This solution would
correspond in the
continuum theory to a negative tension brane.
We investigated the static gravitational
potential and found it realistic when the lattice spacing is
very
large. Last, we also discussed some cosmological solutions.

We can look at our results from two different perspectives.
On one hand, they exemplify the difficulties which arise
upon deconstruction of gravity even at the classical level:
the neglect of the lapse fields leads to spurious solutions,
 while their inclusion only partially solves this problem,
 as we showed in section 3.
On the other hand, they point out interesting directions
in multigravity theories: they allow, as we showed,
simple modifications of gravity at large scales
and brane-like solutions with no branes.
 One  can speculate on the possibility to reproduce
 these results,
in some
more complex multigravity, without the various
drawbacks mentioned in the introduction.
It would also be interesting  to find a solution
corresponding to
the discretization of RS type - gravity localizing - metric
 \cite{RS},
and at the same time avoiding fine
tuning conditions.

\section*{Acknowledgements}
We thank E. Dudas and S. Pokorski for
helpful discussions.

\end{document}